# Boundary Differential Equations and Their Applications to Scattering Problems

Wen Geyi

*Abstract*—In this paper, new boundary differential equations for the two-dimensional exterior scattering problem have been derived. It has been shown that the Helmholtz equation can be reduced to an inhomogeneous Bessel's equation in a body-fitted coordinate system. By imposing the Sommerfeld radiation condition to the general solution of the Bessel's equation, an integro-differential equation, which is equivalent to the original Helmholtz equation, can be obtained. The boundary differential equation can then be established by use of the integration by parts to get rid of the integral in the integro-differential equation for high frequency problems. Numerical examples have been presented to demonstrate the validity of the new boundary differential equations.

*Index Terms*—Absorbing boundary conditions, scattering, boundary differential equations.

## I. INTRODUCTION

It is known that a physical problem can be locally characterized by a differential equation and globally by an integral equation. The former can be numerically solved by a domain method (such as the finite element method and the finite difference method), yielding a sparse matrix; while the latter can be solved by a boundary method (such as the moment method and the boundary element method), yielding a dense matrix. The order of the sparse matrix obtained from the differential equation formulation is usually much higher than the dense matrix from the integral equation formulation.

When the differential equation is defined in a domain of infinite extent, an artificial boundary (truncated boundary) must be introduced to truncate the domain to keep the unknowns finite. To simulate the original field behavior on the truncated boundary without introducing too much distortion, an appropriate boundary condition called absorbing boundary condition must be imposed to diminish the artificial reflections from the truncated boundary. Many research endeavors have been made to devise and improve various absorbing boundary conditions [1]-[19].

The absorbing boundary conditions are basically a relationship or a differential equation relating the field and its normal derivative on the truncated boundary. Kriegsmann, Taflove and Umashankar reported an interesting technique, called on-surface radiation conditions(OSRC), to solve a two-dimensional electromagnetic scattering problem [7], which was then extended and elaborated by many other authors[8]-[19]. Instead of some distance away from the scatterer, the OSRC directly applies the radiation boundary conditions to the surface of the scatterer, which constitutes an Ansatz in the formulation of OSRC. The current distribution can then be obtained analytically for TM incidence or by solving an ordinary differential equation for TE incidence. The above procedure of obtaining the OSRC is a gigantic step as described by Jones, and its validity has never been rigorously proved, and therefore is open to question [8].

The OSRC is essentially a differential equation defined on the surface of the scatterer and it may be conveniently referred to as the boundary differential equation (BDE). The boundary differential equation, when it is solved numerically, costs much less computational time as it results in a sparse matrix while the order of the matrix is much smaller since all the unknowns are confined to the surface of the scatterer.

In this paper, new boundary differential equations for the scattering problem in two-dimensional space will be derived by using a more rigorous approach without resorting to the Ansatz as in the formulation of OSRC. A body-fitted coordinate system will be used [8],[13] in which the Helmholtz equation can be reduced to the well-known inhomogeneous Bessel's equation. By imposing the Sommerfeld radiation condition to the general solution of the Bessel's equation, an integro-differential equation can be obtained, which gives a functional relationship between the field and its normal derivative. For high frequency problems, the integral appearing in the integro-differential equation can be eliminated by use of integration by parts, yielding the boundary differential equations on the surface of the scatterer, which are quite different from the previous ones derived from the OSRC. The numerical results will be presented to demonstrate the applicability and effectiveness of the new boundary differential equations.

## II. HELMHOLTZ EQUATION IN BODY-FITTED COORDINATE SYSTEM

The two-dimensional scalar Helmholtz equation is given by
$$(\nabla^2 + k^2)\phi(\mathbf{r}) = 0, \ \mathbf{r} \in R^2 - \Omega, \qquad (1)$$
where $k$ is the wavenumber, $\Omega$ is a convex domain bounded by $\Gamma$. Let $\tilde{\Gamma}$ be a closed curve enclosing $\Omega$. We assume that $\tilde{\Gamma}$ is parallel to $\Gamma$ in the sense that $\tilde{\Gamma}$ is obtained by a constant offset

W. Geyi is with the College of Electronic and Information Engineering, Nanjing University of Information Science and Technology, Nanjing, 210044, P.R. China, (e-mail: wengeyi@rogers.com).

from $\Gamma$ in the direction of outward normal of $\Gamma$. An arbitrary point $P$ on $\tilde{\Gamma}$ can then be represented by

$$\mathbf{r}(\rho,s) = \mathbf{r}_0(s) + \rho \mathbf{u}_n(s),$$

where $\mathbf{u}_n$ is the unit outward normal of $\Gamma$; $\mathbf{r}_0(s)$ is the local parametric representation of $\Gamma$ in terms of the arc length $s$; $\rho$ is the distance along the normal from the point $\mathbf{r}_0(s)$ to $P$, as shown in Figure 1. It is obvious that the curvilinear coordinate system $(\rho,s)$ is orthogonal and the metric coefficients in the usual notations are given by

$$h_1 \equiv \sqrt{\frac{\partial \mathbf{r}}{\partial s} \cdot \frac{\partial \mathbf{r}}{\partial s}} = h_1^0 [1 + \rho \kappa_0(s)],$$
$$h_2 \equiv \sqrt{\frac{\partial \mathbf{r}}{\partial \rho} \cdot \frac{\partial \mathbf{r}}{\partial \rho}} = 1,$$
(2)

where $h_1^0(s)$ is the value of $h_1$ on $\rho=0$ and is defined by $\sqrt{(\partial \mathbf{r}_0/\partial s)\cdot(\partial \mathbf{r}_0/\partial s)}$; $\kappa_0(s)$ is the curvature of $\Gamma$ at $\rho=0$. In deriving the above equations we have used the relation

$$\frac{d\mathbf{u}_n}{ds} = \kappa_0(s) \frac{d\mathbf{r}_0}{ds}.$$

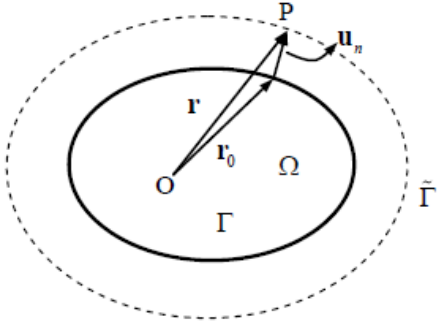

Figure 1 A scatterer in a body-fitted coordinate system

The curvature $\kappa$ of $\tilde{\Gamma}$ at $P$ can thus be expressed in terms of $\kappa_0$ as

$$\kappa(\rho,s) = \frac{\kappa_0(s)}{1+\rho\kappa_0(s)}.$$
(3)

In the curvilinear orthogonal system $(\rho,s)$, we have

$$\nabla^2 = \frac{1}{h_1 h_2}\frac{\partial}{\partial s}\left(\frac{h_2}{h_1}\frac{\partial \phi}{\partial s}\right) + \frac{1}{h_1 h_2}\frac{\partial}{\partial \rho}\left(\frac{h_1}{h_2}\frac{\partial \phi}{\partial \rho}\right)$$
$$= \frac{1}{h_1}\frac{\partial}{\partial s}\left(\frac{1}{h_1}\frac{\partial \phi}{\partial s}\right) + \frac{1}{h_1}\left(h_1\frac{\partial^2 \phi}{\partial \rho^2} + \frac{\partial h_1}{\partial \rho}\frac{\partial \phi}{\partial \rho}\right).$$

Since $s$ is the arc length we have $h_1^0 = 1$, and (1) can be rewritten as

$$\frac{\partial^2 \phi}{\partial \rho^2} + \frac{\kappa_0(s)}{1+\rho\kappa_0(s)}\frac{\partial \phi}{\partial \rho} + k^2\phi = \frac{1}{1+\rho\kappa_0(s)}f(\rho),$$
(4)

where

$$f(\rho) = -\frac{\partial}{\partial s}\left(\frac{1}{1+\rho\kappa_0}\frac{\partial \phi}{\partial s}\right).$$
(5)

If $\kappa_0 \neq 0$ we may let $\xi = (1+\rho\kappa_0)/\kappa_0$ and (4) reduces to

$$\frac{\partial^2 \phi}{\partial \xi^2} + \frac{1}{\xi}\frac{\partial \phi}{\partial \xi} + k^2\phi = \frac{f}{\xi\kappa_0}.$$
(6)

Therefore the Helmholtz equation (1) can be transformed into an inhomogeneous Bessel's equation of zeroth order in the body-fitted coordinate system for an arbitrary convex scatterer.

### III. THE BOUNDARY DIFFERENTIAL EQUATIONS

The general solution of (6) can be expressed as

$$\phi(\xi) = AJ_0(k\xi) + BN_0(k\xi) + \phi_0(\xi),$$
(7)

where $A$ and $B$ are two constants independent of $\xi$; $J_0$ and $N_0$ are the zeroth order Bessel functions of first and second kind respectively; $\phi_0$ is a particular solution of (6) and can be found as follows

$$\phi_0(\xi) = -J_0(k\xi) \int_{1/\kappa_0}^{\xi} \frac{N_0(k\xi)f(\xi)}{\xi\kappa_0 W(J_0,N_0)}d\xi$$
$$+ N_0(k\xi) \int_{1/\kappa_0}^{\xi} \frac{J_0(k\xi)f(\xi)}{\xi\kappa_0 W(J_0,N_0)}d\xi.$$

where $W(J_0,N_0)$ is the Wronskian determinant of the functions $J_0$ and $N_0$. Making use of the following calculation

$$W(J_0,N_0) = k\left[J_0(k\xi)N_0'(k\xi) - J_0'(k\xi)N_0(k\xi)\right]$$
$$= 2/\pi\xi,$$
(8)

Here the prime denotes the derivative with respect to the argument of the Bessel functions. On the boundary $\Gamma$, we have $\xi = 1/\kappa_0$. It follows from (7) that

$$\phi|_\Gamma = AJ_0(\nu) + BN_0(\nu),$$
$$\left.\frac{\partial \phi}{\partial n}\right|_\Gamma = AkJ_0'(\nu) + BkN_0'(\nu).$$
(9)

where $\nu = k/\kappa_0$. From (8) and (9) we obtain

$$A = \frac{\pi\nu}{2k}\left[kN_0'(\nu)\phi|_\Gamma - N_0(\nu)\left.\frac{\partial \phi}{\partial n}\right|_\Gamma\right],$$
$$B = \frac{\pi\nu}{2k}\left[-kJ_0'(\nu)\phi|_\Gamma + J_0(\nu)\left.\frac{\partial \phi}{\partial n}\right|_\Gamma\right].$$
(10)

Thus the general solution (7) can be expressed as

$$\phi(\xi) = \frac{\pi k\nu}{2k}\left[N_0'(\nu)J_0(k\xi) - J_0'(\nu)N_0(k\xi)\right]\phi|_\Gamma$$
$$+ \frac{\pi\nu}{2k}\left[J_0(\nu)N_0(k\xi) - J_0(k\xi)N_0(\nu)\right]\left.\frac{\partial \phi}{\partial n}\right|_\Gamma$$
$$- \frac{\pi}{2\kappa_0}J_0(k\xi)\int_{1/\kappa_0}^{\xi} N_0(k\xi)f(\xi)d\xi$$
$$+ \frac{\pi}{2\kappa_0}N_0(k\xi)\int_{1/\kappa_0}^{\xi} J_0(k\xi)f(\xi)d\xi$$
(11)

and its normal derivative is given by

$$\frac{\partial \phi}{\partial n} = \frac{\pi k v}{2}[N_0'(v)J_0'(k\xi) - J_0'(v)N_0'(k\xi)]\phi\big|_\Gamma$$
$$+ \frac{\pi v}{2}[J_0(v)N_0'(k\xi) - N_0(v)J_0'(k\xi)]\frac{\partial \phi}{\partial n}\bigg|_\Gamma$$
$$- \frac{k\pi}{2\kappa_0}J_0'(k\xi)\int_{1/\kappa_0}^{\xi}N_0(k\xi)f(\xi)d\xi$$
$$+ \frac{k\pi}{2\kappa_0}N_0'(k\xi)\int_{1/\kappa_0}^{\xi}J_0(k\xi)f(\xi)d\xi. \quad (12)$$

When $\rho$ is sufficiently large, $\phi$ must satisfy the radiation condition

$$\lim_{\rho \to \infty}\sqrt{\rho}\left(\frac{\partial \phi}{\partial \rho} + jk\phi\right) = 0. \quad (13)$$

Substituting (11) and (12) into (13) and making use of the asymptotic expressions for the Bessel functions, we obtain

$$kH_1^{(2)}(v)\phi\big|_\Gamma + H_0^{(2)}(v)\frac{\partial \phi}{\partial n}\bigg|_\Gamma$$
$$= \int_0^{+\infty} H_0^{(2)}[v(1+\rho\kappa_0)]\frac{\partial}{\partial s}\left(\frac{1}{1+\rho\kappa_0}\frac{\partial \phi}{\partial s}\right)d\rho. \quad (14)$$

Thus the original Helmholtz equation (1) has been transformed into an integro-differential equation as given by (14), which relates the value of the scalar field $\phi$ on the surface of the scatterer to its normal derivative and is a rigorous expression. Making the substitutions

$$H_0^{(2)}[v(1+\rho\kappa_0)] = h_0(\rho)e^{-jk\rho}, \quad f(\rho) = f_0(\rho)e^{-jk\rho}$$

and using integration by parts, we may find that

$$\int_0^{+\infty} H_0^{(2)}[k(1+\rho\kappa_0)/\kappa_0]f(\rho)d\rho$$
$$= \left[-\frac{j3H_0^{(2)}(v)}{4k} + \frac{H_1^{(2)}(v)}{4k}\right]f(0)$$
$$- \frac{1}{4k^2}H_0^{(2)}(v)\frac{\partial f_0(0)}{\partial \rho}$$
$$- \frac{1}{4k^2}\int_0^{+\infty}e^{-j2k\rho}\frac{\partial^2}{\partial \rho^2}[h_0(\rho)f_0(\rho)]d\rho \quad (15)$$

It should be noted that both $f_0(\rho)$ and $h_0(\rho)$ are slowly varying functions of $\rho$. Ignoring the integral in (15) gives

$$\int_0^{+\infty} H_0^{(2)}[k(1+\rho\kappa_0)/\kappa_0]f(\rho)d\rho$$
$$= \left[\frac{j3H_0^{(2)}(v)}{4k} - \frac{H_1^{(2)}(v)}{4k}\right]\frac{\partial^2 \phi}{\partial s^2}\bigg|_\Gamma - \frac{H_0^{(2)}(v)}{4k^2}\cdot \quad (16)$$
$$\left[\kappa_0'\frac{\partial \phi}{\partial s}\bigg|_\Gamma - \frac{\partial^2}{\partial s^2}\frac{\partial \phi}{\partial n}\bigg|_\Gamma + \kappa_0\frac{\partial^2 \phi}{\partial s^2}\bigg|_\Gamma - jk\frac{\partial^2 \phi}{\partial s^2}\bigg|_\Gamma\right].$$

Substituting this into (14) leads to

$$\frac{\partial \phi}{\partial n}\bigg|_\Gamma = \left(-\frac{j}{k} + \frac{1}{4k\tau} + \frac{\kappa_0}{4k^2}\right)\frac{\partial^2 \phi}{\partial s^2}\bigg|_\Gamma$$
$$+ \frac{\kappa_0'}{4k^2}\frac{\partial \phi}{\partial s}\bigg|_\Gamma - \frac{1}{4k^2}\frac{\partial^2}{\partial s^2}\frac{\partial \phi}{\partial n}\bigg|_\Gamma - \frac{k}{\tau}\phi\big|_\Gamma \quad (17)$$

where $\tau = H_0^{(2)}(v)/H_1^{(2)}(v)$. If the terms higher than $1/k^2$ are ignored in (15), we may obtain

$$\frac{\partial \phi}{\partial n}\bigg|_\Gamma = -\left(\frac{j3}{4k} - \frac{1}{4k\tau}\right)\frac{\partial^2 \phi}{\partial s^2}\bigg|_\Gamma - \frac{k}{\tau}\phi\big|_\Gamma. \quad (18)$$

If the terms higher than $1/k$ are ignored in the above equation, we have

$$\frac{\partial \phi}{\partial n}\bigg|_\Gamma = -\frac{k}{\tau}\phi\big|_\Gamma. \quad (19)$$

Equations (17), (18) and (19) are the boundary differential equations for Helmholtz equation. Note that the second-order derivative along the normal direction in (4) has been removed in the boundary differential equations.

## IV. SCATTERING BY CONDUCTING CYLINDERS

The boundary differential equations (17) and (18) will now be used to study the scattering problems by a convex conducting cylinder. An external field incident upon the cylinder induces a current distribution on its surface, which then produces a scattered field. In order to find the current distribution, one has to solve a Dirichlet boundary value problem for TM (transverse magnetic) incidence, where the scalar field $\phi$ is specified on the surface of the conducting cylinder. For TE (transverse electric) incidence, one has to solve a Neumann boundary value problem, where the normal derivative of the field $\phi$ is assumed to be known on the surface of the conducting cylinder.

### A. TM Incidence

As shown in Figure 2, a TM plane wave with electric field $\mathbf{E}^{in}$ is assumed to be incident upon a uniform conducting cylinder at an angle $\alpha$ with respect to the positive $x$-axis

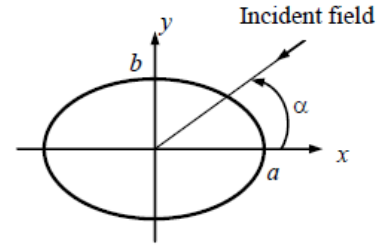

Figure 2 A conducting cylinder illuminated by a plane wave

$$\mathbf{E}^{in} = \mathbf{u}_z E^{in} = \mathbf{u}_z e^{-j\mathbf{k}\cdot\mathbf{r}} = \mathbf{u}_z e^{jk(x\cos\alpha + y\sin\alpha)}.$$

where $\mathbf{u}_z$ is the unit vector along the $z$-direction. The scattered field generated by the induced current distribution on the conducting cylinder is denoted by $\mathbf{E}^s = \mathbf{u}_z E^s$. The total electric field, denoted by $\mathbf{E} = \mathbf{u}_z E$, then satisfies the boundary condition

$$E(\mathbf{r}) = E^{in}(\mathbf{r}) + E^s(\mathbf{r}) = 0, \quad \mathbf{r} \in \Gamma, \quad (20)$$

where $\Gamma$ denotes the boundary of the cross section of the conducting cylinder. We may let $\phi|_\Gamma = E^s$ in (18) to get

$$\left.\frac{\partial E^s}{\partial n}\right|_\Gamma = -\left(\frac{j3}{4k} - \frac{1}{4k\tau}\right)\left.\frac{\partial^2 E^s}{\partial s^2}\right|_\Gamma - \frac{k}{\tau}\left.E^s\right|_\Gamma. \quad (21)$$

The surface current distribution $\mathbf{J}$ on the conducting cylinder is then given by[20]

$$\eta_0 \mathbf{J}(\mathbf{r}) = \mathbf{u}_z \eta_0 J(\mathbf{r}) = \mathbf{u}_z \frac{1}{jk}\left.\frac{\partial E}{\partial n}\right|_\Gamma$$
$$= \mathbf{u}_z \left(\frac{1}{jk}\left.\frac{\partial E^{in}}{\partial n}\right|_\Gamma + \frac{1}{jk}\left.\frac{\partial E^s}{\partial n}\right|_\Gamma\right), \quad (22)$$

where $\eta_0 = 120\pi$ stands for the free space impedance. It follows from (20), (21) and (22) that

$$\eta_0 J = \left(\frac{3}{4k^2} - \frac{1}{j4k^2\tau}\right)\left.\frac{\partial^2 E^{in}}{\partial s^2}\right|_\Gamma + \frac{1}{j\tau}\left.E^{in}\right|_\Gamma + \frac{1}{jk}\left.\frac{\partial E^{in}}{\partial n}\right|_\Gamma. \quad (23)$$

Hence the surface current can be obtained analytically if (18) is used. For higher accuracy, one may use (17). In this case the current satisfies the second-order differential equation

$$\frac{1}{4k^2}\frac{\partial^2 \eta_0 J}{\partial s^2} + \eta_0 J = \left(\frac{1}{k^2} - \frac{1}{j4k^2\tau} - \frac{\kappa_0}{j4k^3}\right)\left.\frac{\partial^2 E^{in}}{\partial s^2}\right|_\Gamma$$
$$-\frac{\kappa_0'}{j4k^3}\left.\frac{\partial E^{in}}{\partial s}\right|_\Gamma + \frac{1}{j4k^3}\left.\frac{\partial^2}{\partial s^2}\frac{\partial E^{in}}{\partial n}\right|_\Gamma + \frac{1}{j\tau}\left.E^{in}\right|_\Gamma + \frac{1}{jk}\left.\frac{\partial E^{in}}{\partial n}\right|_\Gamma.$$

This equation can be solved by using the following one-step iterative scheme

$$\eta_0 J = -\frac{1}{4k^2}\frac{\partial^2 \eta_0 J^{(0)}}{\partial s^2} + \left(\frac{1}{k^2} - \frac{1}{j4k^2\tau} - \frac{\kappa_0}{j4k^3}\right)\left.\frac{\partial^2 E^{in}}{\partial s^2}\right|_\Gamma$$
$$-\frac{\kappa_0'}{j4k^3}\left.\frac{\partial E^{in}}{\partial s}\right|_\Gamma + \frac{1}{j4k^3}\left.\frac{\partial^2}{\partial s^2}\frac{\partial E^{in}}{\partial n}\right|_\Gamma + \frac{1}{j\tau}\left.E^{in}\right|_\Gamma + \frac{1}{jk}\left.\frac{\partial E^{in}}{\partial n}\right|_\Gamma.$$
$$(24)$$

where $\eta_0 J^{(0)}$ is determined by (23). Numerical results indicate that (24) offers very limited improvement compared to (23). For this reason, our numerical examples will be based on (23). In what follows, the starting point of the arc length $s$ is assumed to be on the $x$-axis. Figure 3 shows the amplitudes of the current distributions $\eta_0 J$ on a circular cylinder of unit radius with $k = 6$. The computed results obtained from BDE are compared with those from the method of moment, and both agree very well. Figure 4 shows the current distributions on an elliptical cylinder with semi-major axis $a = 2$ and semi-minor axis $b = 1$ (see Figure 2) for $k = 4$. The elliptical cylinder is illuminated by an incident plane wave from different angles, and the induced current distributions obtained from BDE are in good agreement with those obtained from the moment method. Figure 5 shows the current distributions on a thin strip. The thin strip is approximated by an ellipse with $a = 2, b = 0.01$. It can be seen that the current distribution from BDE is flat on the upper and lower half of the strip, which is similar to the result from physical optics. At the two ends of the strip, the curvature is infinite and the current becomes singular.

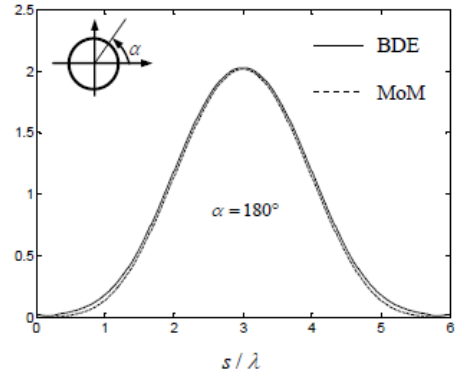

Figure 3 Current on circular cylinder, TM case ($k = 6, \ a = b = 1$)

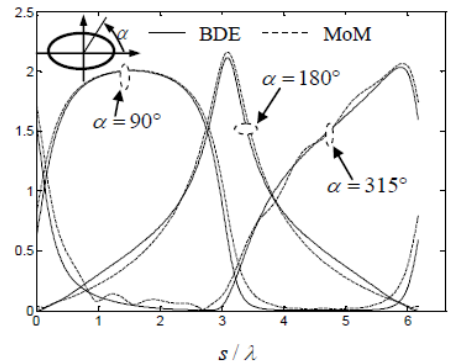

Figure 4 Current on elliptical cylinder, TM case ($k = 4, \ a = 2, \ b = 1$)

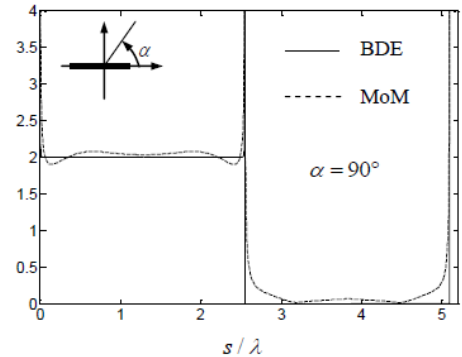

Figure 5 Current on a thin strip, TM case ($k = 4, \ a = 2, \ b = 0.01$)

B. *TE Incidence*

For TE incidence, a plane wave with magnetic field $\mathbf{H}^{in}$ is assumed to be incident upon the uniform conducting cylinder at an angle $\alpha$ with respect to the positive $x$-axis (see Figure 2)

$$\mathbf{H}^{in} = \mathbf{u}_z H^{in} = \mathbf{u}_z e^{-j\mathbf{k}\cdot\mathbf{r}} = \mathbf{u}_z e^{jk(x\cos\alpha + y\sin\alpha)}.$$

The incident field induces a scattered field $\mathbf{H}^s = \mathbf{u}_z H^s$. On the surface of the conducting cylinder, the total magnetic field $\mathbf{H} = \mathbf{u}_z H$ must satisfy the boundary condition

$$\frac{\partial H(\mathbf{r})}{\partial n} = \frac{\partial H^{in}(\mathbf{r})}{\partial n} + \frac{\partial H^s(\mathbf{r})}{\partial n} = 0, \ \mathbf{r} \in \Gamma. \quad (25)$$

Letting $\phi|_\Gamma = H^s$ in (17), (18) and (19), we respectively have

$$\left(-\frac{j3}{4k} + \frac{1}{4k\tau} - \frac{j}{4k} + \frac{\kappa_0}{4k^2}\right)\frac{\partial^2 H^s}{\partial s^2}\bigg|_\Gamma$$
$$+ \frac{\kappa_0'}{4k^2}\frac{\partial H^s}{\partial s}\bigg|_\Gamma - \frac{k}{\tau}H^s\big|_\Gamma = \frac{\partial H^s}{\partial n}\bigg|_\Gamma + \frac{1}{4k^2}\frac{\partial^2}{\partial s^2}\frac{\partial H^s}{\partial n}\bigg|_\Gamma, \quad (26)$$

$$\left(j3k - \frac{k}{\tau}\right)\frac{\partial^2 H^s}{\partial s^2}\bigg|_\Gamma + \frac{4k^3}{\tau}H^s\big|_\Gamma = -4k^2\frac{\partial H^s}{\partial n}\bigg|_\Gamma, \quad (27)$$

and

$$\frac{\partial H^s}{\partial n}\bigg|_\Gamma = -\frac{k}{\tau}H^s\big|_\Gamma. \quad (28)$$

The surface electric current is then given by

$$\mathbf{J}(\mathbf{r}) = \mathbf{u}_\varphi J(\mathbf{r}) = -\mathbf{u}_\varphi \left[H^{in}(\mathbf{r}) + H^s(\mathbf{r})\right]. \quad (29)$$

The current distributions corresponding to (26), (27), and (28) respectively satisfy the following equations

$$J = a(s)\frac{\partial^2 J}{\partial s^2} + b(s)\frac{\partial J}{\partial s} + c(s), \quad (30)$$

$$J = g(s)\frac{\partial^2 J}{\partial s^2} + f(s), \quad (31)$$

$$J = -\frac{\tau}{k}\frac{\partial H^{in}}{\partial n}\bigg|_\Gamma - H^{in}\big|_\Gamma, \quad (32)$$

where

$$a(s) = -\frac{j\tau}{k^2} + \frac{1}{4k^2} + \frac{\kappa_0 \tau}{4k^3}, \ b(s) = \frac{\kappa_0' \tau}{4k^3},$$

$$c(s) = a(s)\frac{\partial^2 H^{in}}{\partial s^2}\bigg|_\Gamma + b(s)\frac{\partial H^{in}}{\partial s}\bigg|_\Gamma - H^{in}\big|_\Gamma$$
$$- \frac{\tau}{k}\frac{\partial H^{in}}{\partial n}\bigg|_\Gamma - \frac{\tau}{4k^3}\frac{\partial^2}{\partial s^2}\frac{\partial H^{in}}{\partial n}\bigg|_\Gamma,$$

$$g(s) = -\frac{j3\tau}{4k^2} + \frac{1}{4k^2},$$

$$f(s) = g(s)\frac{\partial^2 H^{in}}{\partial s^2}\bigg|_\Gamma - H^{in}\big|_\Gamma - \frac{\tau}{k}\frac{\partial H^{in}}{\partial n}\bigg|_\Gamma.$$

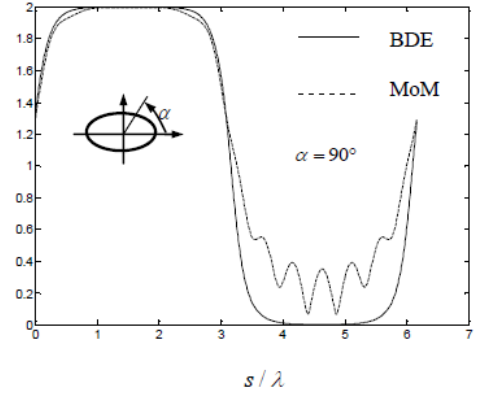

Figure 6 Current on elliptical cylinder, TE case
($k = 4$, $a = 2$, $b = 1$)

Similarly the one-step iterative solutions for the current distributions can be obtained by introducing (32) into the right-hand side of (30) and (31). The following numerical calculations are based on (31). Figure 6 shows the current distribution on the elliptical cylinder for $k = 4$. It can be seen that the current distribution on the shadowed region from BDE is a bit off from the moment solution, and is not as accurate as the TM case. The accuracy of the BDE solutions improves as frequency increases. Figure 7 shows the current distribution for the same elliptical cylinder with $k = 50$. It can be seen that the BDE result improves and approaches to the moment solution.

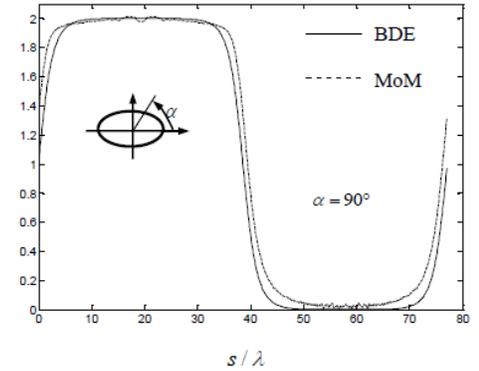

Figure 7 Current on elliptical cylinder, TE case
($k = 50$, $a = 2$, $b = 1$)

V. CONCLUSION

In this paper, we have shown that the two-dimensional Helmholtz equation defined in an exterior region can be transformed into an integro-differential equation in a body-fitted coordinate system for a convex scatterer, which can then be used to derive the boundary differential equations in a rigorous manner. The solutions of the boundary differential equations can be found analytically, and have been demonstrated to be valid and effective for high frequency scattering problems.